# ESG Signaling on Wall Street in the AI Era

*Qionghua Chu*[1]

**Abstract**

I identify a new signaling channel in ESG research by empirically examining whether environmental, social, and governance (ESG) investing remains valuable as large institutional investors increasingly shift toward artificial intelligence (AI). Using winsorized ESG scores of S&P 500 firms from Yahoo Finance and controlling for market value of equity, I conduct cross-sectional regressions to test the signaling mechanism. I demonstrate that Environmental, Social, Governance, and composite ESG scores strongly and positively signal higher debt-to-total-capital ratio[2], both individually and in various combinations. My findings contribute to the growing literature on ESG investing, offering economically meaningful signaling channel with implications for long-term portfolio management amid the rise of AI.

[1] Nanyang Technological University, Nanyang Business School, 52 Nanyang Ave, Singapore 639798. Email: chuq0002@e.ntu.edu.sg.

[2] Debt-to-total-capital ratio is calculated as book value of debt divided by the sum of book value of debt and market value of equity, with the sum commonly referred as total capital.

# 1. Introduction

As artificial intelligence (AI) gains rising significance and becomes the new holy grail, large institutional investors gradually reduce both the size and percentage of their assets under management invested in environmental (E), social (S), and governance (G) – collectively referred to as the ESG theme. As an epitome, BlackRock, the world's largest asset manager, is no longer in climate group as Wall Street tones down on impact investing (Kerber, 2025). As ESG skepticism looms, large institutional investors put more focus on investments in AI (Sen & Mandl, 2025) in aspects such as Large Language Models, Machine Learning, and Deep Reinforcement Learning. This draws a key question: Is ESG investing no longer worth it? I deem the empirical evidence to prove the opposite. ESG provides a good signal on debt-to-total-capital ratio (DTCR).

From a theoretical perspective, my rationale for why ESG scores should signal DTCR is as follows. Similar to Miller and Rock (1985)'s dividend signaling channel, I find the positive association between ESG and DTCR to be consistent with a signaling mechanism. Firms with advanced ESG practices use higher DTCR to credibly convey their lower risk profile to capital providers. Robust ESG frameworks enhance reputation and stakeholder trust, reducing perceived risk and thereby increasing debt capacity, as reflected in DTCR.

Empirically, I investigate if ESG is still worthwhile by examining if it affects the S&P 500 firms' DTCR by conducting a cross-sectional analysis of whether and how S&P 500 firms' individual and any combinations of E, S, G, and total ESG scores indicated on Yahoo Finance associate with the DTCR, with market value of equity (MVE) as the control variable. My single-main-explanatory-variable models use E, S, G, and total ESG scores individually. As for my multi-main-explanatory-variable models, besides the quadruple-main-explanatory-variable model of E, S, G, and total ESG scores as main explanatory variables, the double- and triple-main-explanatory-variable models combine all the possible sets among E, S, G, and total ESG scores as main explanatory variables. To ensure robustness, I have winsorized all five independent variables – total ESG, E, S, and G scores as well as MVE – at the $1^{st}$ and $99^{th}$ percentiles to account for sensitivity to outliers.

For empirical findings, I find a novel signaling channel, advancing the theory of signaling by documenting the ESG effect through the DTCR ratio. I show that with and without MVE considered, individual total ESG, E, and S scores, as well as E and S and E and G scores two-factor combinations strongly and positively signal higher DTCR. G score signals higher DTCR only when MVE is controlled. As for S and G scores combination, only S score strongly and positively signals higher DTCR with MVE considered. On three-factor combinations, when combined with S and G scores, total ESG score signals higher DTCR both with and without MVE considered; when coupled with E and S scores, total ESG score signals higher DTCR with MVE controlled. For the combination of E, S, and G scores, with MVE considered, E and G scores signal higher DTCR, while E score also signals higher DTCR without MVE considered. Total ESG, E, and G scores do not signal higher DTCR with or without MVE controlled. On the four-factor combination of ESG, E, S, and G scores, I do not observe any signaling effect.

On unique contributions, my innovative findings contribute to several strands of the finance literature that examine how ESG performance influences the cost of capital, fund flows, and firm value, and is affected by factors such as tax policy changes, proxy voting, supply chain dynamics, and shareholderism. Firstly, studies so far present that ESG divesture strategies do not affect the cost of capital (Berk & van Binsbergen, 2024), ESG ratings showcase only firms' nonfinancial impact and have major limitations (Larcker et al., 2022), and the lack of sustainability results in net fund outflows (Hartzmark & Sussman, 2019). In comparison, I indicate that ESG scores positively signal higher DTCR. Moreover, whereas Ginglinger and Moreau (2023) demonstrate that greater climate risk reduces leverage, my findings focus on how environmental aspect of ESG could be a positive signal for firms' capability to borrow. Furthermore, while extant studies enlighten that corporate social responsibility (CSR) performance enhances following tax cuts (Chang et al., 2025), corporate clients unilaterally apply CSR effect to suppliers (Dai et al., 2021), and well-governed firms with fewer agency problems to have a positive relation between CSR and shareholder value (Ferrell et al., 2016), my findings focus on how the social facet affects DTCR. Lastly, I manifest how governance affects DTCR, while current studies examine how directors' personal values and cultural harmony affect their strategies towards shareholders and stakeholders (Adams & Licht, 2025), blockholders' abstention might affect corporate governance (Bar-Isaac & Shapiro, 2020), corporate governance affects firm value and operating performance (Bebchuk, et al., 2013), green activism influences corporate governance via strategic proxy voting (Jin and Noe, 2024), and large amounts of short-term debt impair corporate governance (Voss, 2025).

As for real-world applications as contributions, as the cycle of investing turns back to the ESG theme in future, the ESG signaling effect via the DTCR ratio will enable more effective asset allocation. I demonstrate portfolio management implications for portfolio managers' and hedge fund managers' asset allocation strategies given firms' E, S, G, and total ESG scores.

The remainder of my paper is structured as follows. Section 2 delineates data and methodology and my summary statistics. Section 3 characterizes my main results and presents robustness checks for concerns and possible explanations. Section 4 demonstrates effective asset management strategies. Section 5 concludes the findings, discusses broader implications, and widens the scope for future research.

## 2. Data, Hypothesis, and Methodology

### 2.1. Data Source and Robustness Check

On source, I use high-quality data from Yahoo Finance[3] with a sample size of 503 firms in S&P 500 index as of 5th September 2025. In Table 1 Panels A and B, I provide summary statistics to give an overview of features of data covered.

For comprehensive analysis, besides analyzing how total ESG scores might affect DTCR, I analyze how E, S, and G scores respectively and any combinations might affect DTCR.

On robustness check, I conduct a cross-sectional analysis on both single- and multi-variable regressions to have control variables.

### 2.2. ESG Signaling Hypotheses and Evaluation via Single- and Multi-Variable Models

ESG signaling with ESG scores could be achieved via two main ways – ESG or its component scores alone and various combinations. I form two hypotheses. To examine my hypotheses, I conduct cross-sectional regressions. With $\beta_0$ as the intercept coefficient, $\beta_1$, $\beta_2$, $\beta_3$, and $\beta_4$ are coefficients for total ESG, E, S, and G scores correspondingly, and $\beta_5$ is coefficient for the control variable MVE. $Total\ ESG_{i,t}$, $E_{i,t}$, $S_{i,t}$, and $G_{i,t}$ represent total ESG, E, S, and G scores; $MVE_{i,t}$ is market value of equity.

---

[3] Yahoo Finance provides scores on E, S, G, and ESG based on firms' public report, rating agencies – such as MSCI and Sustainalytics, and industry standards.

**Hypothesis 1 (H1) (ESG Signaling Effect for One-Score Model)**: For individual scores, the total ESG score provides an overview of firms' sustainability performance. Not only do top executives have strong incentives to maintain superb ESG practices, but they are also highly motivated to use higher DTCR to strengthen their firms' risk management profiles. As E, S, and G scores are components of the total ESG score, I expect them to individually deliver similar signaling effects.

I investigate H1 via regressions (1) and (2). Regression (1) assesses how the total ESG score affect DTCR, while regression (2) adds in the control variable MVE. To apply the equation across other main-explanatory variables, as an example, for $\beta_1 \times Total\ ESG_{i,t}$ in (1) and (2), any other product of an independent variable and its coefficient, such as $\beta_2 \times E_{i,t}$, could replace it.

$$DTCR_{i,t} = \beta_0 + \beta_1 \times Total\ ESG_{i,t} + \varepsilon_{i,t}, \qquad (1)$$

$$DTCR_{i,t} = \beta_0 + \beta_1 \times Total\ ESG_{i,t} + \beta_5 \times MVE_{i,t} + \varepsilon_{i,t}. \qquad (2)$$

**Hypothesis 2 (H2) (ESG Signaling Effect for Two- and Three-Score Models)**: On combination of scores, while a four-factor combination of total ESG, E, S, and G scores risks multicollinearity, two- and three-factor combinations of total ESG, E, S, and G scores could have some indicators to provide similar ESG signaling effects as the individual scores. When top management teams view different ESG combinations, they might find one or a few scores to be more effective to signal to investors the firms' superior performance and hence, capability to undertake more debts and avoid defaults. This signals higher DTCR. In turn, the senior executives, including C-suite, are incentivized to sustain exemplary ESG practices.

To assess H2, I form regressions (3) to (8). Regression (3) shows double-main-explanatory-variable model of how total ESG and E scores affect DTCR, with regression (4) controlling for MVE. For applications on other combinations, on $\beta_1 \times Total\ ESG_{i,t} + \beta_2 \times E_{i,t}$ in (3) and (4), any possible combination of the sum of two products of main-explanatory variables and their coefficients such as $\beta_2 \times E_{i,t} + \beta_3 \times S_{i,t}$ could replace it.

$$DTCR_{i,t} = \beta_0 + \beta_1 \times Total\ ESG_{i,t} + \beta_2 \times E_{i,t} + \varepsilon_{i,t}, \qquad (3)$$

$$DTCR_{i,t} = \beta_0 + \beta_1 \times Total\ ESG_{i,t} + \beta_2 \times E_{i,t} + \beta_5 \times MVE_{i,t} + \varepsilon_{i,t}. \quad (4)$$

On triple-main-explanatory-variable model, I demonstrate it via regression (5) and apply MVE as the control in regression (6). While both regressions assess how total ESG, E, and S scores affect DTCR, to apply for other combinations, for instance, the combination of E, S, and G scores, $\beta_1 \times Total\ ESG_{i,t} + \beta_2 \times E_{i,t} + \beta_3 \times S_{i,t}$ could be replaced with $\beta_2 \times E_{i,t} + \beta_3 \times S_{i,t} + \beta_4 \times G_{i,t}$.

$$DTCR_{i,t} = \beta_0 + \beta_1 \times Total\ ESG_{i,t} + \beta_2 \times E_{i,t} + \beta_3 \times S_{i,t} + \varepsilon_{i,t}, \quad (5)$$

$$DTCR_{i,t} = \beta_0 + \beta_1 \times Total\ ESG_{i,t} + \beta_2 \times E_{i,t} + \beta_3 \times S_{i,t} + \beta_5 \times MVE_{i,t} + \varepsilon_{i,t}. \quad (6)$$

As for regression (7), I show quadruple-main-explanatory-variable model of how total ESG, E, S, and G scores affect DTCR, with MVE as a control in regression (8).

$$DTCR_{i,t} = \beta_0 + \beta_1 \times Total\ ESG_{i,t} + \beta_2 \times E_{i,t} + \beta_3 \times S_{i,t} + \beta_4 \times G_{i,t} + \varepsilon_{i,t}, \quad (7)$$

$$DTCR_{i,t} = \beta_0 + \beta_1 \times Total\ ESG_{i,t} + \beta_2 \times E_{i,t} + \beta_3 \times S_{i,t} + \beta_4 \times G_{i,t} + \beta_5 \times MVE_{i,t} + \varepsilon_{i,t}. \quad (8)$$

## 3. Main Results

### 3.1. Key Empirical Contributions

My empirical analysis gives innovative evidence of the signaling power of ESG scores on DTCR. Besides, my findings manifestly distinguish the signaling effects of individual and any combinations of E, S, G, and total ESG scores and provides significant insights into DTCR behavior. In Tables 2 to 5, I show findings and contributions, which are presented via a heatmap in Figure 1. My findings are robust to alternative treatments of outliers without winsorization, with results available in Appendices A.2 and A.3 Panels A to C.

### 3.2. Individual E, S, G, and Total ESG Scores Positively Signal Higher DTCR

*3.2.1. Robustness of Results and Economic Interpretation*

In Table 2, I display findings of signaling regression of single-variable model. I have tested all 4 scores with and without the control variable. I show that E, S, and total ESG scores signal higher DTCR both with and without control variables, while G score does so only with MVE controlled. This evidence is consistent with H1.

Firms' outstanding environmental and social practices are more externally visible, which could affect firms' overall ESG practices to enable firms to deliver reliable images to investors to continue providing capitals. As governance practices, while affecting firm's share prices and MVE, are internal and relatively less transparent (Hermalin & Weisbach, 2007) to draw investors' attentions (Iliev et al., 2021), firms might view these conducts to be less likely to affect investors' confidence on their DTCR capability and focus less on keeping great measures.

*3.2.2. Magnitude of Impact*

I elucidate magnitude of impact of $\beta_1$ values in single-main-explanatory-variable models. To quantify impact, as per Table 2, for a firm with a rise in total capital of $1 billion, an increase in score of 1 for total ESG signals a rise in book value of debt by $3.58 million. For proportional economic relevance, an uplift of 1 total ESG score indicates a 0.358% increment in book value of debt for a 1% growth in total capital.

**3.3. E, S, G, and Total ESG Scores Positively Signal Higher DTCR via Various Combinations**

*3.3.1. Robustness of Results and Economic Interpretation*

As for Table 3 Panels A to C, I demonstrate results of signaling regression of multi-main-explanatory-variable models. In Panel A of Table 3, on two-factor combinations, for both with and without control variables, I demonstrate that for the combination of total ESG with E, S, or G scores, only total ESG score signals higher DTCR, while E and S as well as E and G scores signal higher DTCR. As for S and G scores combination, only S score signals higher DTCR with the control variable in place. The results are consistent with H2.

Moreover, in Table 3 Panel B, for three-factor combinations, total ESG signals higher DTCR when it combines with S and G scores, and when it combines with E and S scores, it only signals higher DTCR with the control variable in place. For the combination of E, S, and G scores, E score signals higher DTCR both with and without the control variable, and G score only does so with control variable in the regression. Total ESG, E, and G combination does not signal higher DTCR. These support H2.

As for the four-factor combination of total ESG, E, S, and G scores in Panel C of Table 3, I do not observe any signaling effect. This accords with the interpretation of multicollinearity underlying H2.

Overall, the total ESG score dominates in combinations to most effectively signal firms' borrowing capacity and DTCR, while governance provides the weakest signal due to lower investor attention (Iliev et al., 2021) and transparency (Hermalin & Weisbach, 2007), reducing managerial incentives to maintain or enhance it.

*3.3.2. Magnitude of Impact*

I delineate magnitude of impact of $\beta_1$, $\beta_2$, $\beta_3$, and $\beta_4$ values via multi-main-explanatory-variable models. For economic relevance, as an example, based on Table 3 Panel A, with MVE in consideration, the rise in score of 1 in a double-variable model of S and G scores signals an increase in book value of debt of $4.02 million for a firm with a growth in total capital of $1 billion. Percentage-wise, the increase in S and G scores of 1 reveals a 0.402% rise in book value of debt when the total capital raises by 1%.

## 4. Portfolio Management Implications

### 4.1. Total ESG, E, S, and G Scores Strategies

Financially, outstanding total ESG, E, S, and G scores present better firm images is intuitive, as good practices instill reliability into investors to justify firms' higher DTCR ability. Portfolio managers should invest in firms which could maintain or improve their total ESG, E, S, and G scores, as firms' management-level executives are keen to do so if they have ability to keep or enhance their DTCR. However, if portfolio managers notice fall in scores in total ESG, E, S, or G, they should put the firms' securities on a watch list to decide if the fall is temporary or permanent to hold or sell respectively to maintain great returns for portfolios. Alternatively, for hedge fund managers, the fall in scores in total ESG, E, S, or G is a great signal to short the firms' equities until the rise in scores become imminent.

Moreover, as climate risk is key for institutional investors (Krueger et al., 2020), firms might deem the social governance or factor to be important only when investors also focus on the environmental aspect. Portfolio and hedge fund managers should appraise the social or governance factor in tandem with the environmental aspect. Portfolio and hedge fund managers should not keep or increase current holdings if only firms' S or G score rises; they should only do so when E score increases simultaneously.

Furthermore, when it comes to considering all three factors – E, S, and G scores – together, portfolio and hedge fund managers should expect only E and G scores to signal higher DTCR. If portfolio managers focus on total ESG score, they should observe it together with E and S or S and G scores combinations. As for the four-factor combination of total ESG, E, S, and G scores, portfolio and hedge fund managers do not have to focus on it.

## 5. Contributions

### 5.1. Theoretically Extend Signaling Theory to ESG

My findings extend the signaling effect of dividends on firms' financial health to the public (Miller and Rock, 1985) to ESG. I demonstrate that consistent with a signaling channel, ESG scores are positively associated with higher DTCR, even in combinations. Senior management has an information advantage regarding firm's fund sufficiency, and their decision to maintain or enhance – rather than weaken – ESG practices signals the firm's borrowing capacity to investors.

### 5.2. Empirically Provide New Evidence across S&P 500

Empirically, with the voluntary reporting of ESG practices by firms and addition of ESG scores by institutions such as Sustainalytics and MSCI, my results reveal new angles to view S&P 500 to assess firms' ESG scores to determine their borrowing capability. In addition, given that top executives are motivated to sustain or enhance their firms' ESG practices to signal optimism to debtholders, this provides positive externalities to society.

### 5.3. Practically Explore Portfolio Allocation Strategies

In real world, portfolio managers could view the same, rise, or fall of E, S, G, and total ESG scores as a good signal to hold, buy, or sell S&P 500 firms' stocks and bonds. As for hedge fund managers, they could view the fall of the scores as a good signal to short and the rise to long S&P 500 firms' equities and fixed income securities, with the same scores as a safety net to hold.

## 6. Conclusion

In my paper, I form pronounced and novel findings. Distinct from existing research, I indicate that total ESG, E, S, and G scores strongly and positively signal higher DTCR individually and in multiple combinations. This highlights significance of ESG for institutional investors while AI crowds it out. ESG and AI could co-exist as bellwethers in investment themes. Moreover, when ESG turns to become the main investment theme again in future, this ESG signaling effect will rise in importance. For robustness, I validate my findings by reducing sensitivity to outliers through winsorization at the $1^{st}$ and $99^{th}$ percentiles. For real-life applications, I evaluate how portfolio managers and hedge fund managers could adjust their strategies based on the same, rising, or falling total ESG, E, S, and G scores.

On broader implications of my findings, first, policymakers could encourage firms to increase corporate governance transparency to enable investors to make more informed decisions on firms' DTCR capability. Besides, regulators could enhance standards on how firms report their ESG practices so that total ESG, E, S, and G scores assessed by institutions such as Yahoo Finance could be more accurate to facilitate investors to make informed decisions. Finally, law enforcement agencies could consider making more ESG practices mandatory to report to enable investors to have more clarity to determine DTCR.

As empirical studies are rarely flawless, this study entails certain limitations. One plausible concern is that ESG ratings might be lagging indicators to reflect firms' actual ESG performance. However, firms have incentives to maintain strong ESG profiles to signal robust borrowing capability. Furthermore, the vast number of ESG raters allow for specialization (Azarmsa & Shapiro, 2025), making individual E, S, and G scores informative on firms' credit strength. Therefore, albeit lagging, ESG scores sustain signaling effect on firms' debt capacity.

For future research, I deem it worthwhile to investigate firstly, how E, S, G, and total ESG affect DTCR on a time-series basis for a minimum of 10 years. Furthermore, promising research could go beyond the S&P 500 or US equities market to explore if the same findings hold true in other markets, such as Singapore. Last but not least, exploring how AI could enhance firms' E, S, G, and ESG scores to signal higher DTCR would be thought-provoking.

# Figure 1
## Heatmap for ESG Signaling Effect

Figure 1 presents the signaling effect of total ESG, E, S, and G scores on DTCR, with total ESG, E, S, and G refer to total ESG, Social, and Governance scores respectively. I indicate green and red for positive and negative E, S, G, and total ESG coefficients correspondingly. Sets (1) to (15) indicate single- and multi- main-explanatory-variable models tested, with regression results with and without MVE as the control variable indicated. My data source is Yahoo Finance and sample period is as of 5$^{th}$ September 2025. ***, **, and * to denote significance at the 1%, 5%, and 10% levels (two-sided) in sequence.

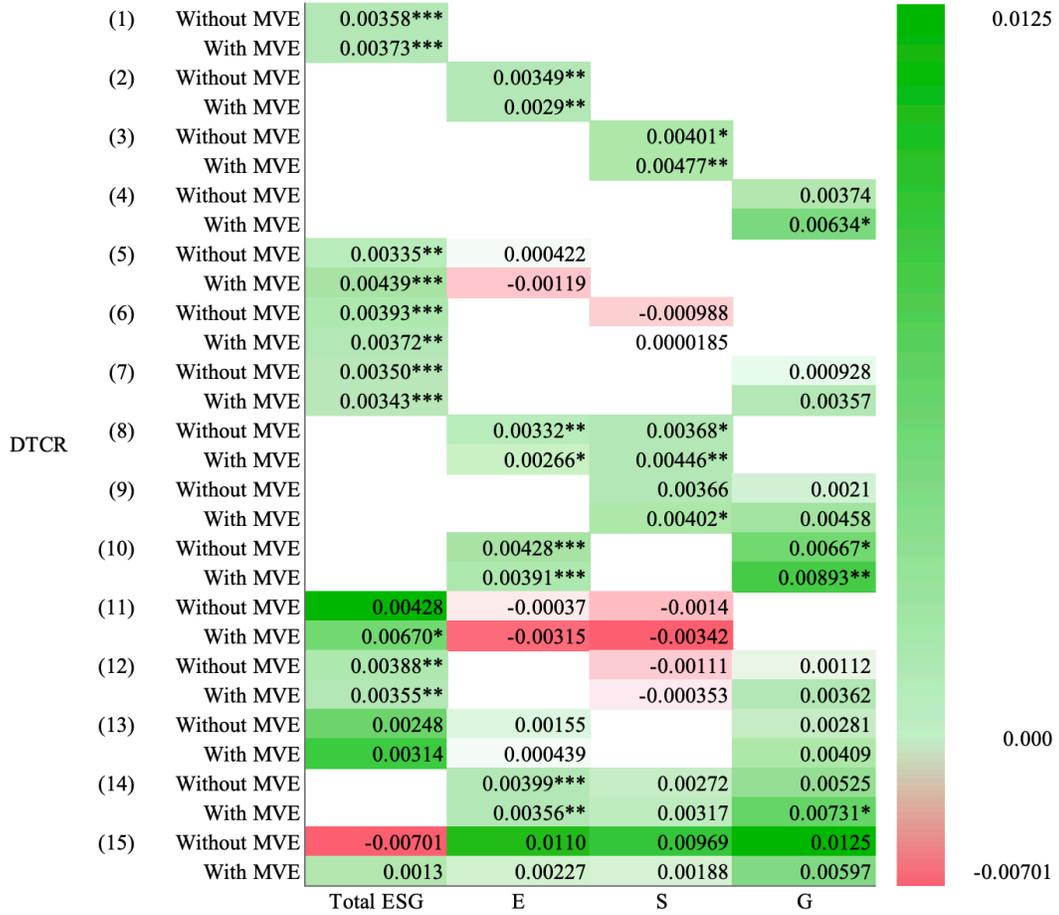

| | | | Total ESG | E | S | G |
|---|---|---|---|---|---|---|
| DTCR | (1) | Without MVE | 0.00358*** | | | |
| | | With MVE | 0.00373*** | | | |
| | (2) | Without MVE | | 0.00349** | | |
| | | With MVE | | 0.0029** | | |
| | (3) | Without MVE | | | 0.00401* | |
| | | With MVE | | | 0.00477** | |
| | (4) | Without MVE | | | | 0.00374 |
| | | With MVE | | | | 0.00634* |
| | (5) | Without MVE | 0.00335** | 0.000422 | | |
| | | With MVE | 0.00439*** | -0.00119 | | |
| | (6) | Without MVE | 0.00393*** | | -0.000988 | |
| | | With MVE | 0.00372** | | 0.0000185 | |
| | (7) | Without MVE | 0.00350*** | | | 0.000928 |
| | | With MVE | 0.00343*** | | | 0.00357 |
| | (8) | Without MVE | | 0.00332** | 0.00368* | |
| | | With MVE | | 0.00266* | 0.00446** | |
| | (9) | Without MVE | | | 0.00366 | 0.0021 |
| | | With MVE | | | 0.00402* | 0.00458 |
| | (10) | Without MVE | | 0.00428*** | | 0.00667* |
| | | With MVE | | 0.00391*** | | 0.00893** |
| | (11) | Without MVE | 0.00428 | -0.00037 | -0.0014 | |
| | | With MVE | 0.00670* | -0.00315 | -0.00342 | |
| | (12) | Without MVE | 0.00388** | | -0.00111 | 0.00112 |
| | | With MVE | 0.00355** | | -0.000353 | 0.00362 |
| | (13) | Without MVE | 0.00248 | 0.00155 | | 0.00281 |
| | | With MVE | 0.00314 | 0.000439 | | 0.00409 |
| | (14) | Without MVE | | 0.00399*** | 0.00272 | 0.00525 |
| | | With MVE | | 0.00356** | 0.00317 | 0.00731* |
| | (15) | Without MVE | -0.00701 | 0.0110 | 0.00969 | 0.0125 |
| | | With MVE | 0.0013 | 0.00227 | 0.00188 | 0.00597 |

## Table 1

## Summary Statistics for S&P 500 Firms' DTCR and ESG Scores

Table 1 presents summary statistics for the S&P 500 firms' DTCR, total ESG, E, S, and G scores, and MVE. DTCR refers to book value of debt divided by the sum of book value of debt and market value of equity, with the sum denoting total capital. My data source is Yahoo Finance. My sample period is as of 5$^{th}$ September 2025.

**Panel A. Descriptive Statistics on Dependent Variable**

|  | Value |
|---|---|
| Average DTCR | 0.216 |
| Median DTCR | 0.177 |
| Maximum DTCR | 0.816 |
| Minimum DTCR | 0.000 |

**Panel B. Descriptive Statistics on Independent Variables**

|  | Value |
|---|---|
| Average Total ESG Score | 20.9 |
| Median Total ESG Score | 20.2 |
| Maximum Total ESG Score before Winsorization | 45.1 |
| Minimum Total ESG Score before Winsorization | 7.28 |
| 99$^{th}$ Percentile Total ESG Score before Winsorization | 38.3 |
| 1$^{st}$ Percentile Total ESG Score before Winsorization | 9.00 |
| Maximum Total ESG Score after Winsorization | 38.3 |
| Minimum Total ESG Score after Winsorization | 9.00 |
| Average E Score | 6.29 |
| Median E Score | 4.54 |
| Maximum E Score before Winsorization | 25.5 |
| Minimum E Score before Winsorization | 0.0400 |
| 99$^{th}$ Percentile E Score before Winsorization | 20.9 |
| 1$^{st}$ Percentile E Score before Winsorization | 0.0794 |
| Maximum E Score after Winsorization | 20.9 |
| Minimum E Score after Winsorization | 0.0794 |
| Average S Score | 9.18 |
| Median S Score | 9.20 |
| Maximum S Score before Winsorization | 21.6 |
| Minimum S Score before Winsorization | 0.890 |
| 99$^{th}$ Percentile S Score before Winsorization | 17.7 |
| 1$^{st}$ Percentile S Score before Winsorization | 1.47 |
| Maximum S Score after Winsorization | 17.7 |
| Minimum S Score after Winsorization | 1.47 |
| Average G Score | 5.41 |
| Median G Score | 4.91 |
| Maximum G Score before Winsorization | 19.2 |
| Minimum G Score before Winsorization | 1.68 |
| 99$^{th}$ Percentile G Score before Winsorization | 17.7 |
| 1$^{st}$ Percentile G Score before Winsorization | 1.47 |
| Maximum G Score after Winsorization | 17.7 |
| Minimum G Score after Winsorization | 1.47 |
| Average MVE | $1.12 \times 10^{11}$ |
| Median MVE | $3.79 \times 10^{10}$ |
| Maximum MVE before Winsorization | $4.07 \times 10^{12}$ |
| Minimum MVE before Winsorization | $7.93 \times 10^{9}$ |
| 99$^{th}$ Percentile MVE after Winsorization | $1.58 \times 10^{12}$ |
| 1$^{st}$ Percentile MVE after Winsorization | $8.43 \times 10^{9}$ |
| Maximum MVE after Winsorization | $1.58 \times 10^{12}$ |
| Minimum MVE after Winsorization | $8.43 \times 10^{9}$ |

## Table 2

## Regression Results for ESG Signaling via Single-Main-Explanatory-Variable Model

Table 2 shows regression results on the signaling power of ESG scores with winsorization via the single-main-explanatory-variable model. My data source is the same as that in Table 1. Sets (1) to (4) have the first and second sets of rows for without and with control variables respectively. DTCR represents book value of debt divided by the sum of book value of debt and market value of equity, with the sum denoting total capital. $β_0$ is intercept; $β_1$, $β_2$, $β_3$, and $β_4$ are coefficient on total ESG, E, S, and G scores respectively, with $β_5$ as the control variable coefficient for MVE. Standard errors in parentheses. Total ESG, E, S, and G scores have been winsorized at the 1st and 99th percentiles. My sample period is as of 5th September 2025. I use ***, **, and * to denote significance at the 1%, 5%, and 10% levels (two-sided) correspondingly.

| DTCR | Constant | $β_1$ | $β_2$ | $β_3$ | $β_4$ | $β_5$ | Adjusted $R^2$ |
|---|---|---|---|---|---|---|---|
| (1) | 0.142*** | 0.00358*** | | | | | 0.0184 |
| | (0.0245) | (0.00112) | | | | | |
| | 0.154*** | 0.00373*** | | | | $-1.55×10^{-13}$*** | 0.0546 |
| | (0.0242) | (0.00110) | | | | $(3.47×10^{-14})$ | |
| (2) | 0.195*** | | 0.00349** | | | | 0.00973 |
| | (0.0112) | | (0.00144) | | | | |
| | 0.212*** | | 0.00290** | | | $-1.44×10^{-13}$*** | 0.0405 |
| | (0.0123) | | (0.00143) | | | $(3.52×10^{-14})$ | |
| (3) | 0.180*** | | | 0.00401* | | | 0.00511 |
| | (0.0209) | | | (0.00213) | | | |
| | 0.188** | | | 0.00477** | | $-1.58×10^{-13}$*** | 0.0425 |
| | (0.0206) | | | (0.00210) | | $(3.50×10^{-14})$ | |
| (4) | 0.196*** | | | | 0.00374 | | 0.000321 |
| | (0.0201) | | | | (0.00347) | | |
| | 0.198*** | | | | 0.00634* | $-1.62×10^{-13}$*** | 0.0391 |
| | (0.0197) | | | | (0.00345) | $(3.55×10^{-14})$ | |

# Table 3

# Regression Results for ESG Signaling via Multi-Main-Explanatory-Variable Models

Table 3 demonstrates regression results on the signaling power of ESG scores with winsorization via the double-, triple-, and quadruple-main-explanatory-variable model. My data source is the same as that in Table 1. Sets (5) to (15) have the first and second sets of rows for without and with control variables respectively. DTCR represents book value of debt divided by the sum of book value of debt and market value of equity, with the sum denoting total capital. $\beta_0$ is intercept; $\beta_1$, $\beta_2$, $\beta_3$, and $\beta_4$ are coefficient on total ESG, E, S, and G scores respectively, with $\beta_5$ as the control variable coefficient for MVE. Standard errors in parentheses. Total ESG, E, S, and G scores have been winsorized at the 1st and 99th percentiles. My sample period is as of 5th September 2025. I use ***, **, and * to denote significance at the 1%, 5%, and 10% levels (two-sided) correspondingly.

**Panel A**. ESG Signaling via Double-Main-Explanatory-Variable Model

| DTCR | Constant | $\beta_1$ | $\beta_2$ | $\beta_3$ | $\beta_4$ | $\beta_5$ | Adjusted $R^2$ |
|---|---|---|---|---|---|---|---|
| (5) | 0.144*** | 0.00335** | 0.000422 | | | | 0.0165 |
| | (0.0268) | (0.00160) | (0.00205) | | | | |
| | 0.148*** | 0.00439*** | -0.00119 | | | $-1.59\times10^{-13}$*** | 0.0533 |
| | (0.0264) | (0.00159) | (0.00205) | | | ($3.53\times10^{-14}$) | |
| (6) | 0.143*** | 0.00393*** | | -0.000988 | | | 0.0166 |
| | (0.0250) | (0.00151) | | (0.00286) | | | |
| | 0.154*** | 0.00372** | | 0.0000185 | | $-1.55\times10^{-13}$*** | 0.0527 |
| | (0.0247) | (0.00148) | | (0.00282) | | ($3.49\times10^{-14}$) | |
| (7) | 0.138*** | 0.00350*** | | | 0.000928 | | 0.0165 |
| | (0.0277) | (0.00116) | | | (0.00357) | | |
| | 0.141*** | 0.00343*** | | | 0.00357 | $-1.61\times10^{-13}$*** | 0.0546 |
| | (0.0272) | (0.00114) | | | (0.00355) | ($3.52\times10^{-14}$) | |
| (8) | 0.162*** | | 0.00332** | 0.00368* | | | 0.0137 |
| | (0.0222) | | (0.00144) | (0.00213) | | | |
| | 0.174*** | | 0.00266* | 0.00446** | | $-1.51\times10^{-13}$*** | 0.0473 |
| | (0.0220) | | (0.00143) | (0.00210) | | ($3.52\times10^{-14}$) | |
| (9) | 0.172*** | | | 0.00366 | 0.00210 | | 0.00378 |
| | (0.0251) | | | (0.00222) | (0.00361) | | |
| | 0.171*** | | | 0.00402* | 0.00458 | $-1.65\times10^{-13}$*** | 0.0438 |
| | (0.02458) | | | (0.00218) | (0.00357) | ($3.54\times10^{-14}$) | |
| (10) | 0.154*** | | 0.00428*** | | 0.00667* | | 0.0146 |
| | (0.0250) | | (0.00150) | | (0.00360) | | |
| | 0.159*** | | 0.00391*** | | 0.00893** | $-1.57\times10^{-13}$*** | 0.0507 |
| | (0.0245) | | (0.00148) | | (0.00357) | ($3.53\times10^{-14}$) | |

**Panel B. ESG Signaling via Triple-Main-Explanatory-Variable Model**

| DTCR | Constant | $\beta_1$ | $\beta_2$ | $\beta_3$ | $\beta_4$ | $\beta_5$ | Adjusted $R^2$ |
|---|---|---|---|---|---|---|---|
| (11) | 0.142*** | 0.00428 | -0.000370 | -0.00140 | | | 0.0146 |
| | (0.0275) | (0.00353) | (0.00336) | (0.00469) | | | |
| | 0.143*** | 0.00670* | -0.00315 | -0.00342 | | $-1.61\times10^{-13}$*** | 0.0525 |
| | (0.0270) | (0.00350) | (0.00335) | (0.00462) | | $(3.55\times10^{-14})$ | |
| (12) | 0.140*** | 0.00388** | | -0.00111 | 0.00112 | | 0.0148 |
| | (0.0279) | (0.00152) | | (0.00289) | (0.00361) | | |
| | 0.142*** | 0.00355** | | -0.000353 | 0.00362 | $-1.60\times10^{-13}$*** | 0.0527 |
| | (0.0274) | (0.00149) | | (0.00284) | (0.00358) | $(3.53\times10^{-14})$ | |
| (13) | 0.140*** | 0.00248 | 0.00155 | | 0.00281 | | 0.0151 |
| | (0.0279) | (0.00222) | (0.00287) | | (0.00499) | | |
| | 0.142*** | 0.00314 | 0.000439 | | 0.00409 | $-1.60\times10^{-13}$*** | 0.0527 |
| | (0.0273) | (0.00218) | (0.00283) | | (0.00490) | $(3.54\times10^{-14})$ | |
| (14) | 0.138*** | | 0.00399*** | 0.00272 | 0.00525 | | 0.0155 |
| | (0.0280) | | (0.00152) | (0.00224) | (0.00378) | | |
| | 0.141*** | | 0.00356** | 0.00317 | 0.00731* | $-1.59\times10^{-13}$*** | 0.0528 |
| | (0.0275) | | (0.00149) | (0.00220) | (0.00374) | $(3.53\times10^{-14})$ | |

**Panel C. ESG Signaling via Quadruple-Main-Explanatory-Variable Model**

| DTCR | Constant | $\beta_1$ | $\beta_2$ | $\beta_3$ | $\beta_4$ | $\beta_5$ | Adjusted $R^2$ |
|---|---|---|---|---|---|---|---|
| (15) | 0.138*** | -0.00701 | 0.0110 | 0.00969 | 0.0125 | | 0.0140 |
| | (0.0280) | (0.0139) | (0.0139) | (0.0140) | (0.0149) | | |
| | 0.141*** | 0.00130 | 0.00227 | 0.00188 | 0.00597 | $-1.60\times10^{-13}$*** | 0.0508 |
| | (0.0275) | (0.0137) | (0.0137) | (0.0138) | (0.0147) | $(3.57\times10^{-14})$ | |

## Appendix A.1: Abbreviations for Key Variables

In this table, I indicate how main variables in this paper are abbreviated.

| Variable | Abbreviations |
|---|---|
| Environmental, Social, and Governance | ESG |
| Environmental | E |
| Social | S |
| Governance | G |
| Debt-to-Total-Capital Ratio | DTCR |
| Market Value of Equity | MVE |

**Appendix A.2: Regression Results for ESG Signaling without Winsorization via Single-Main-Explanatory-Variable Model**

In this table, I demonstrate regression results on the signaling power of ESG scores without winsorization via the single-main-explanatory-variable model. My data source is the same as that in Table 1. Sets (1) to (4) have the first and second sets of rows for without and with control variables correspondingly. Debt-to-total-capital ratio (DTCR) represents book value of debt divided by the sum of book value of debt and market value of equity, with the sum denoting total capital. $\beta_0$ is intercept; $\beta_1$, $\beta_2$, $\beta_3$, and $\beta_4$ are coefficient on total ESG, E, S, and G scores respectively, with $\beta_5$ as the control variable coefficient for MVE. Standard errors in parentheses. Total ESG, E, S, and G scores have been not winsorized. My sample period is as of 5$^{th}$ September 2025. I use ***, **, and * to denote significance at the 1%, 5%, and 10% levels (two-sided) respectively.

| DTCR | Constant | $\beta_1$ | $\beta_2$ | $\beta_3$ | $\beta_4$ | $\beta_5$ | $R^2$ Adjusted |
|---|---|---|---|---|---|---|---|
| (1) | 0.145*** | 0.00341*** | | | | | 0.0171 |
| | (0.0241) | (0.0011) | | | | | |
| | 0.155*** | 0.00339*** | | | | $-7.93 \times 10^{-14}$*** | 0.0427 |
| | (0.024) | (0.00109) | | | | ($2.11* \times 10^{-14}$) | |
| (2) | 0.195*** | | 0.00335** | | | | 0.00911 |
| | (0.0116) | | (0.00142) | | | | |
| | 0.207*** | | 0.00294** | | | $-7.93 \times 10^{-14}$*** | 0.0323 |
| | (0.0119) | | (0.00141) | | | ($2.13* \times 10^{-14}$) | |
| (3) | 0.183*** | | | 0.00367* | | | 0.00414 |
| | (0.0206) | | | (0.00210) | | | |
| | 0.190*** | | | 0.00391* | | $-8.10 \times 10^{-14}$*** | 0.0308 |
| | (0.0204) | | | (0.00207) | | ($2.12* \times 10^{-14}$) | |
| (4) | 0.199*** | | | | 0.00324 | | -0.0000816 |
| | (0.0194) | | | | (0.00330) | | |
| | 0.200*** | | | | 0.00481 | $-8.36 \times 10^{-14}$*** | 0.0280 |
| | (0.0191) | | | | (0.00328) | ($2.14* \times 10^{-14}$) | |

**Appendix A.3**: **Regression Results for ESG Signaling without Winsorization via Multi-Main-Explanatory-Variable Models**

In this table, I show regression results on the signaling power of ESG scores without winsorization via the double-, triple-, and quadruple-main-explanatory-variable model. My data source is the same as that in Table 1. Sets (5) to (10) have the first and second sets of rows for without and with control variables correspondingly. Debt-to-total-capital ratio (DTCR) represents book value of debt divided by the sum of book value of debt and market value of equity, with the sum denoting total capital. $β_0$ is intercept; $β_1$, $β_2$, $β_3$, and $β_4$ are coefficient on total ESG, E, S, and G scores respectively, with $β_5$ as the control variable coefficient for MVE. Standard errors in parentheses. Total ESG, E, S, and G scores have been not winsorized. My sample period is as of 5$^{th}$ September 2025. I use ***, **, and * to denote significance at the 1%, 5%, and 10% levels (two-sided) respectively.

**Panel A**. ESG Signaling without winsorization via Double-Main-Explanatory-Variable Model

| DTCR | Constant | $β_1$ | $β_2$ | $β_3$ | $β_4$ | $β_5$ | $R^2$ Adjusted |
|---|---|---|---|---|---|---|---|
| (5) | 0.147*** | 0.00319** | 0.000392 | | | | 0.0152 |
| | (0.02656) | (0.00159) | (0.00204) | | | | |
| | 0.152*** | 0.00365** | -0.000463 | | | -7.99×10$^{-14}$*** | 0.0408 |
| | (0.0262) | (0.00157) | (0.00203) | | | (2.12*×10$^{-14}$) | |
| (6) | 0.147*** | 0.00382** | | -0.00116 | | | 0.0154 |
| | (0.0247) | (0.00148) | | (0.00280) | | | |
| | 0.156*** | 0.00362** | | -0.000676 | | -7.92×10$^{-14}$*** | 0.0408 |
| | (0.0245) | (0.00146) | | (0.00277) | | (2.11*×10$^{-14}$) | |
| (7) | 0.143*** | 0.00335*** | | | 0.000629 | | 0.0152 |
| | (0.0270) | (0.00114) | | | (0.00340) | | |
| | 0.147*** | 0.00318*** | | | 0.00229 | -8.12×10$^{-14}$*** | 0.0416 |
| | (0.0270) | (0.00113) | | | (0.00338) | (2.13*×10$^{-14}$) | |
| (8) | 0.166*** | | 0.00320** | 0.00335 | | | 0.0122 |
| | (0.0219) | | (0.00142) | (0.00209) | | | |
| | 0.174*** | | 0.00276* | 0.00363* | | -7.75×10$^{-14}$*** | 0.0363 |
| | (0.0218) | | (0.00141) | (0.00207) | | (2.12*×10$^{-14}$) | |
| (9) | 0.176*** | | | 0.00336 | 0.00186 | | 0.00272 |
| | (0.0246) | | | (0.00217) | (0.00342) | | |
| | 0.176*** | | | 0.00335 | 0.00344 | -8.36×10$^{-14}$*** | 0.0308 |
| | (0.0242) | | | (0.00214) | (0.00339) | (2.14*×10$^{-14}$) | |
| (10) | 0.160*** | | 0.00404*** | | 0.00579* | | 0.0129 |
| | (0.0240) | | (0.00148) | | (0.00341) | | |
| | 0.163*** | | 0.00376** | | 0.00714** | -8.08×10$^{-14}$*** | 0.0390 |
| | (0.0237) | | (0.00146) | | (0.00339) | (2.13*×10$^{-14}$) | |

**Panel B**. ESG Signaling without winsorization via Triple-Main-Explanatory-Variable Model

| DTCR | Constant | $\beta_1$ | $\beta_2$ | $\beta_3$ | $\beta_4$ | $\beta_5$ | $R^2$ Adjusted |
|---|---|---|---|---|---|---|---|
| (11) | 0.145*** | 0.00459 | -0.000808 | -0.00205 | | | 0.0136 |
|  | (0.0273) | (0.00357) | (0.00342) | (0.00470) | | | |
|  | 0.148*** | 0.00591* | -0.00242 | -0.00332 | | $-8.10\times10^{-14}$*** | 0.0399 |
|  | (0.0270) | (0.00354) | (0.00340) | (0.00464) | | ($2.13*\times10^{-14}$) | |
| (12) | 0.145*** | 0.00378** | | -0.00124 | 0.000811 | | 0.0136 |
|  | (0.0273) | (0.00149) | | (0.00283) | (0.00342) | | |
|  | 0.148*** | 0.00349** | | -0.000906 | 0.00242 | $-8.10\times10^{-14}$*** | 0.0399 |
|  | (0.0270) | (0.00147) | | (0.00279) | (0.00340) | ($2.13*\times10^{-14}$) | |
| (13) | 0.145*** | 0.00253 | 0.00125 | | 0.00206 | | 0.0136 |
|  | (0.0273) | (0.00219) | (0.00283) | | (0.00470) | | |
|  | 0.148*** | 0.00259 | 0.000908 | | 0.00333 | $-8.10\times10^{-14}$*** | 0.0399 |
|  | (0.0270) | (0.00216) | (0.00279) | | (0.00464) | ($2.13*\times10^{-14}$) | |
| (14) | 0.145*** | | 0.00378** | 0.00253 | 0.00459 | | 0.0136 |
|  | (0.0273) | | (0.00149) | (0.00219) | (0.00357) | | |
|  | 0.148*** | | 0.00349** | 0.00259 | 0.00592* | $-8.10\times10^{-14}$*** | 0.0399 |
|  | (0.0270) | | (0.00147) | (0.00216) | (0.00354) | ($2.13*\times10^{-14}$) | |

**Panel C**. ESG Signaling without winsorization via Quadruple-Main-Explanatory-Variable Model

| DTCR | Constant | $\beta_1$ | $\beta_2$ | $\beta_3$ | $\beta_4$ | $\beta_5$ | $R^2$ Adjusted |
|---|---|---|---|---|---|---|---|
| (15) | 0.145*** | -0.336 | 0.340 | 0.339 | 0.341 | | 0.0117 |
|  | (0.0274) | (1.16) | (1.16) | (1.16) | (1.16) | | |
|  | 0.148*** | -0.328 | 0.331 | 0.330 | 0.334 | $-8.10\times10^{-14}$*** | 0.0381 |
|  | (0.0270) | (1.15) | (1.15) | (1.15) | (1.15) | ($2.13*\times10^{-14}$) | |